\documentclass[lettersize,journal]{IEEEtran}
\usepackage{amsmath,amsfonts}
\usepackage{algorithmic}
\usepackage{algorithm}
\usepackage{array}
\usepackage[caption=false,font=normalsize,labelfont=sf,textfont=sf]{subfig}
\usepackage{textcomp}
\usepackage{stfloats}
\usepackage{url}
\usepackage{verbatim}
\usepackage{graphicx}
\usepackage{cite}
\begin{document}

\title{A Non-planar ReBCO Test Coil with 3D-printed Aluminum Support Structure for the EPOS Stellarator}

\author{Paul Huslage, Tristan Schuler, Pedro F. Gil, Vitali Brack, Dylan Schmeling, Diego A. R. Orona, Elisabeth von Schoenberg, Timo Thun, J. Smoniewski and E. V. Stenson
\thanks{Manuscript received xxxx; revised xxxx.}}

\markboth{Transactions of Applied Superconductivity}%
{Shell \MakeLowercase{\textit{et al.}}: A Sample Article Using IEEEtran.cls for IEEE Journals}


\maketitle

\begin{abstract}
We report on the test of a small scale, non-planar coil using non-insulated ReBCO tape wound on a 3D-printed aluminum support structure. A 3D-scan of the winding frame which was printed out of AlSi10Mg using selective laser melting, showed peak manufacturing deviations of 0.3\,mm. We tested the coil with 21 turns of 3\,mm wide tape cooled both by liquid nitrogen and with a cryocooler. We achieved a central field strength of up to 21\,mT which agrees with the prediction suggesting that we manufactured the coil without defects. The peak field was reached at a supply current of 120\,A. The current leads showed a contact resistance of $\left(2.25\pm0.13\right)\mu\Omega$. The discharge time $L/R$ was found to be 0.63\,s in liquid nitrogen and 2.92\,s when cooled by the cryocooler. From this we estimate a winding pack temperature of 41\,K. 
\end{abstract}


\section{Introduction}
\IEEEPARstart{R}{eBCO} high-temperature superconductors are an attractive prospective technology for fusion reactors \cite{hartwig_sparc_2024, endrizzi_physics_2023}, plasma science experiments \cite{michael_development_2017}, and other fields of research (e.g, quantum materials research \cite{li_rebco_2024} neutron spectroscopy \cite{winn_flexible_2022} or axion research \cite{kim_design_2020}) due to the higher achievable magnetic field strength and/or operating temperature compared to low temperature superconductors like NiTi and Ni$_3$Sn.  
One of these examples is EPOS (a tabletop-sized stellarator, which aims to confine an electron positron pair plasma in a 2\,T optimized magnetic field \cite{stoneking_new_2020,huslage_strain_2024}. The engineering design concept includes non-planar ReBCO coils conductively cooled with cold heads inside a simply connected, cylindrical vacuum chamber. 

Non-insulated (NI) ReBCO coils do not use any form of insulation between individual turns of ReBCO tape. This allows for current and heat load sharing between layers. They offer a high degree of thermal stability and passive quench protection which makes them a promising candidate for high field magnets \cite{hartwig_sparc_2024, li_rebco_2024, winn_flexible_2022, kim_design_2020}. 

Due to their brittle, ceramic nature and anisotropic bending properties, the manufacturing of non-planar coils from ReBCO is a significant engineering challenge. Special care must be taken to avoid damage from torsion and bending induced strains. This can be done by optimizing the orientation of the orientation of the tape stack \cite{huslage_strain_2024, paz-soldan_non-planar_2020} Another approach is to wind the ReBCO tapes into a cable, which leads to an isotropic bending behavior at the cost of much higher bending radii \cite{hartwig_viper_2020, goldacker_roebel_2014}. Non-planar, low field ReBCO coils were constructed using each of those approaches \cite{huslage_winding_2024, riva_development_2023}. 

However, none of the previous projects demonstrated the operation of a non-planar coil using ReBCO tape at the small scale ($d\approx 0.2$\,m) needed for the EPOS stellarator. The small scale leads to high torsional and bending strains which may prevent the use of ReBCO for the coils of EPOS unless special care is taken of coil optimization with regards to the strain on the superconducting tape. 

\begin{figure}[bt]
    \centering
    \includegraphics[width=\columnwidth]{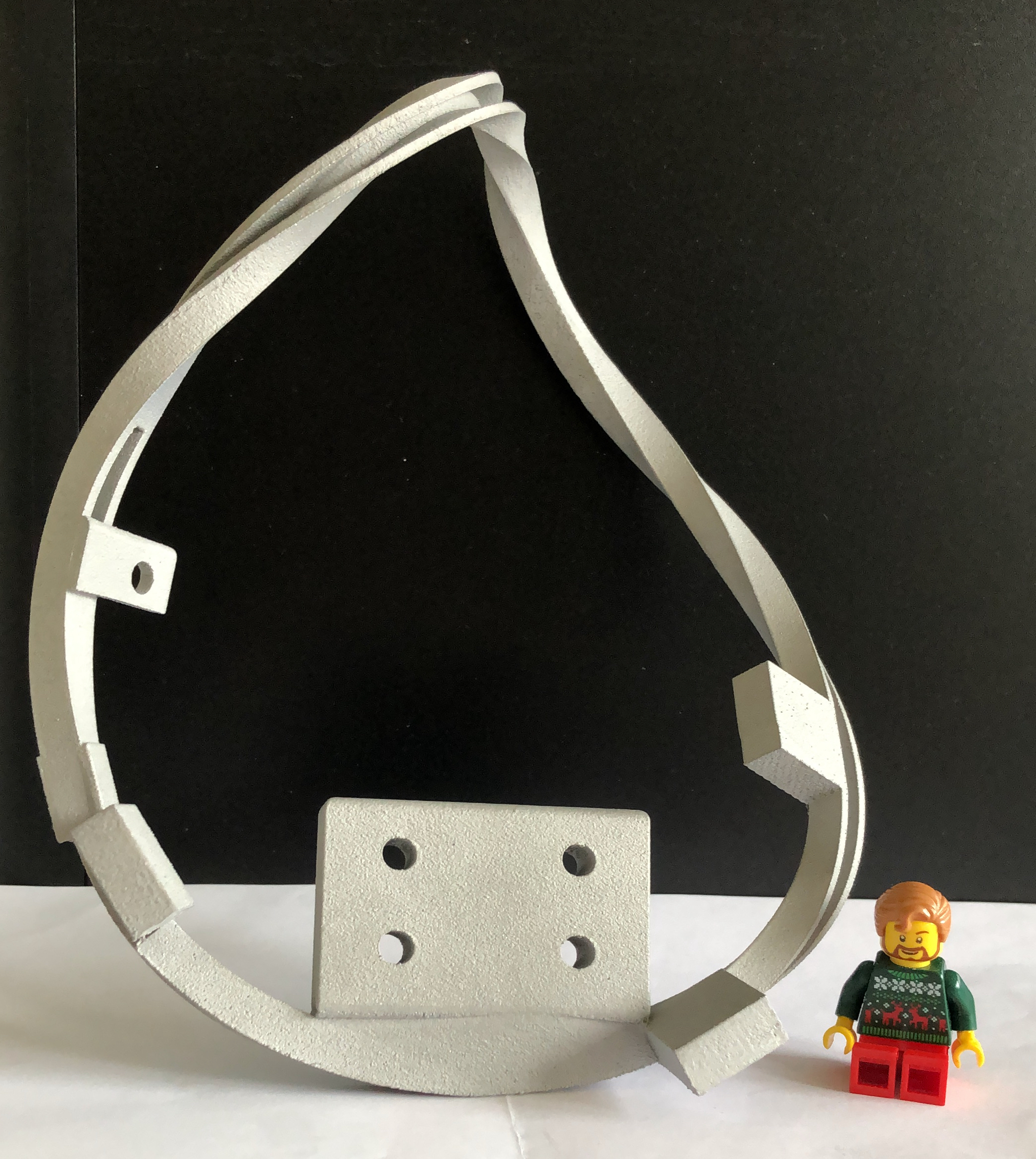}
    \caption{3D printed, strain-optimized support structure for the non-planar test coil made from AlSi10Mg by 3D activation. The three blocks are meant to support radial current leads. The plate with 4 holes clamps the a mouunt on the cryocooler. The amount of non-planarity on such a small coil approaches the strain limit of 0.2\% used in the EPOS optimization. LEGO portrait of the first author for scale.}
    \label{fig:supoort-frame}
\end{figure}

This necessitates the development of several test coils before the design and assembly of the EPOS stellarator in order to inform optimization and design choices. Apart from strain on the superconducting tape, several challenges in cooling and energizing the coils arise. The coils need to be cooled conductively with only a limited amount of cooling power ($\leq100$\,W) at 20\,K operating temperature while balancing the heat loads from black body radiation off the chamber walls as well as conduction and Ohmic losses at a comparatively high supply current ($\approx 350$\,A) from the copper current leads.

Furthermore, stellarator coils have notoriously tight (and sometimes deal-breaking) engineering tolerances  \cite{rise_experiences_2009, strykowsky_engineering_2009}. 
They are also very complicated, which raises question of additive manufacturing (AM) / 3D printing) as a potentially attractive solution. It is common belief, that "Complexity comes for free" in AM processes. The layer-wise build up of parts seemingly provides a mechanism whereby the manufacturing time and therefore cost of the part does not depend on its complexity. But the irregular shapes of non-planar coil frames pose a unique challenge to these typical tolerances as well as the accuracy of the manufacturing processes. Today's metal 3D-printing processes have manufacturing tolerances of several 100s of $\mu$m, whereas the manufacturing tolerance of the EPOS device will be $\approx 1$\,mm. A careful investigation of the manufacturing process choosen for EPOS is required since significant deviations from the optimized magnetic field lead to an increase of particle losses.

In this paper, we describe an EPOS test coil that verified winding angle optimization at the relevant (10-cm-scale) coil sizes, using a 3D-printed aluminum winding frame.  

In section \ref{sec:strain_opt}, we describe the optimization of the coil shape and the manufacturing of the winding frame. We show results of energizing the coil both being submerged in liquid nitrogen and cooled by a cryocooler in section \ref{sec:measurements}. We obtain contact resistance, discharge time and the magnetic field. We compare those to analytical calculations of the expected magnetic field. Furthermore, we calculate an estimate of the winding pack temperature from the difference in resistance of the copper surround between liquid nitrogen and cold head cooling. In section \ref{sec:summary}, we summarize the results and describe how the coil presented in this paper informs the EPOS engineering design and give an outlook on both future test coils as well as the construction of the device. 

\section{Strain Optimized 3D-printed, non-planar winding frame}\label{sec:strain_opt}
The shape of the current path and winding angle of the tape were optimized using the SIMSOPT stellarator optimization framework \cite{landreman_simsopt_2021}. To avoid damage to the ReBCO tape through strain from torsion and hard-way bending, we kept the total strain $\epsilon$ below 0.2\% through an optimized orientation of the winding pack. Details on the strain optimization can be found in \cite{huslage_strain_2024}. This was then turned into a CAD design for the test coil winding frame, with trench dimensions of 3.2\,mm width and 5\,mm height to allow for a stack of 3\,mm wide ReBCO tape from THEVA and a flat plate for connecting to the cold head. Several radial extending blocks were added to support 12\,mm current leads

The strain-optimized coil frame was printed by 3D-activation in the AlSi10Mg aluminum alloy using selective laser melting. Figure \ref{fig:supoort-frame} shows the 3D-printed coil frame. To address the suitability for our specific use case, measurements of the 3D-printed aluminum coil-winding frames were conducted with an Absolute Scanner AS1 from Hexagon Metrology.
These were compared to the CAD design using the measurement software's best fit option. The largest deviation is $\approx0.3\,$mm and is located in the most non-planar, most complicated area of the coil. This is in line with the specifications of the supplier. A detailed analysis of manufacturing deviations from multiple test coils and different manufacturing approaches will be presented in future work.


%

\section{Liquid Nitrogen and cryocooler Tests} \label{sec:measurements}
\begin{figure}
    \centering
    \includegraphics[width=\columnwidth]{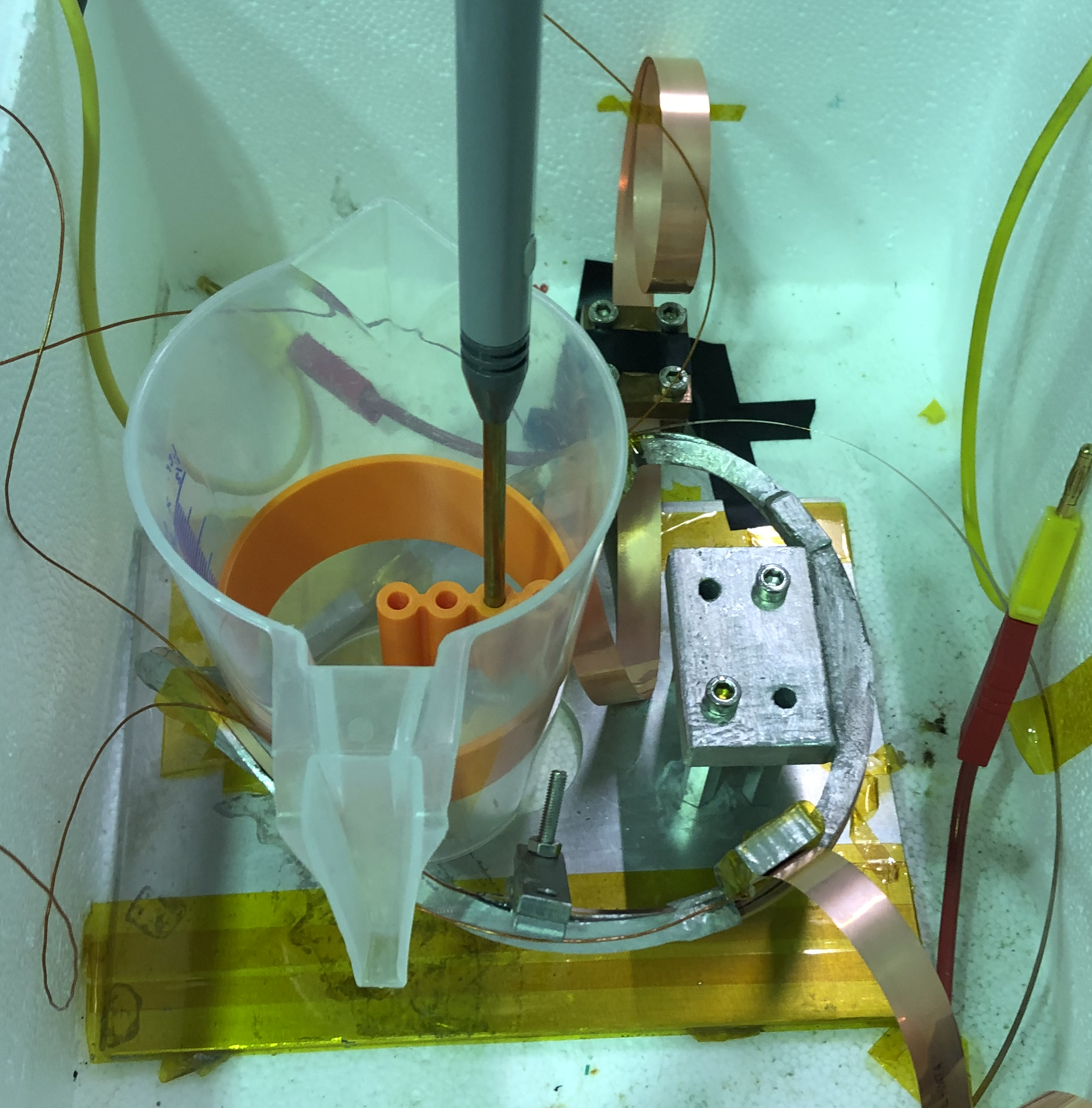}
    \caption{Setup for testing the coil in liquid nitrogen. Current is fed via radially soldered 12\,mm ReBCO strips. The axial (name) Hall probe is mounted in a cup that is filled with nuts to avoid buoyancy. Voltage is measured across the winding pack as well as the solder contacts.}
    \label{fig:ln2_setup}
\end{figure}

For testing at 77 K, we submerged the coil in liquid nitrogen inside a styrofoam box.  The liquid nitrogen test setup is depicted in figure \ref{fig:ln2_setup}.
Voltage taps were soldered onto the 12\,mm wide ReBCO current leads, which were clamped to copper blocks, in order to feed the current into the coil. Further voltage taps were soldered directly onto the first and last winding of the 3\,mm winding pack. 

The magnetic field of the coil was measured using the AP002 axial hall probe which was positioned close to the center of the coil using a plastic cup with a 3D printed adapter. The cup was held in place by being filled with (non-magnetic) A4 stainless steel nuts to avoid movement by buoyancy in liquid nitrogen.  

\begin{figure}[!t]
\centering
\includegraphics[width=\columnwidth]{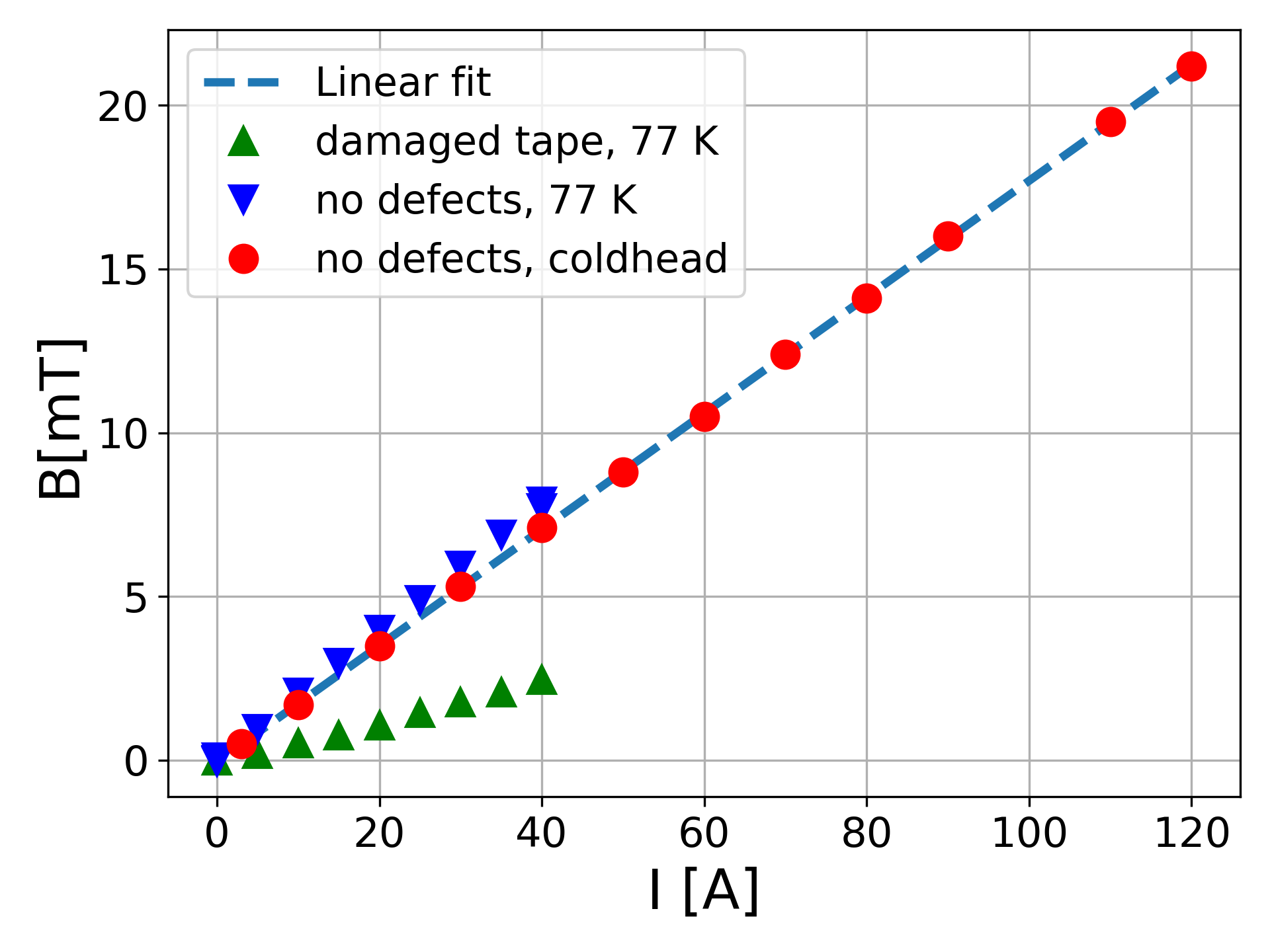}%
\caption{Magnetic field for different supply current measured both for liquid nitrogen and cold head cooling. Both measurements taking from the rewound coil follow the analytical calculation from the Biot-Savart law. The damaged coil shows significantly less magnetic field due to defects along the superconducting winding pack.}
\label{fig:bfield}
\end{figure}

For the first winding test, 13 turns of 3\,mm wide HTS tape on 50\,$\mu$m substrate with 10\,$\mu$m copper surround were coated with BiSnAg from ChipQuik\textregistered solder paste, wound onto the frame with a prototype winding machine , then baked at 160°C for 20\,min. Leads made of 12-mm ReBCO tape were radially soldered into the winding pack.  The coil was energized up to 40\,A. The current was limited by the power supply. The magnetic field of the soldered coil showed poor performance and reached only 2.46\,mT (see figure \ref{fig:bfield}) which falls short of the expected value. This indicates that defects were introduced to the ceramic ReBCO layer of the tape during winding or soldering. The winding frame showed a high level of surface roughness which made it harder to wind tape around it and may very well have introduced defects. Only after extensive polishing with a hand tool was the tape not getting stuck between the walls of the frame. However, it is worth noting that the coil still showed stable operation in liquid nitrogen without any signs of quenching due to the robust nature of NI coils.

\begin{figure}
    \centering
    \includegraphics[width=\columnwidth]{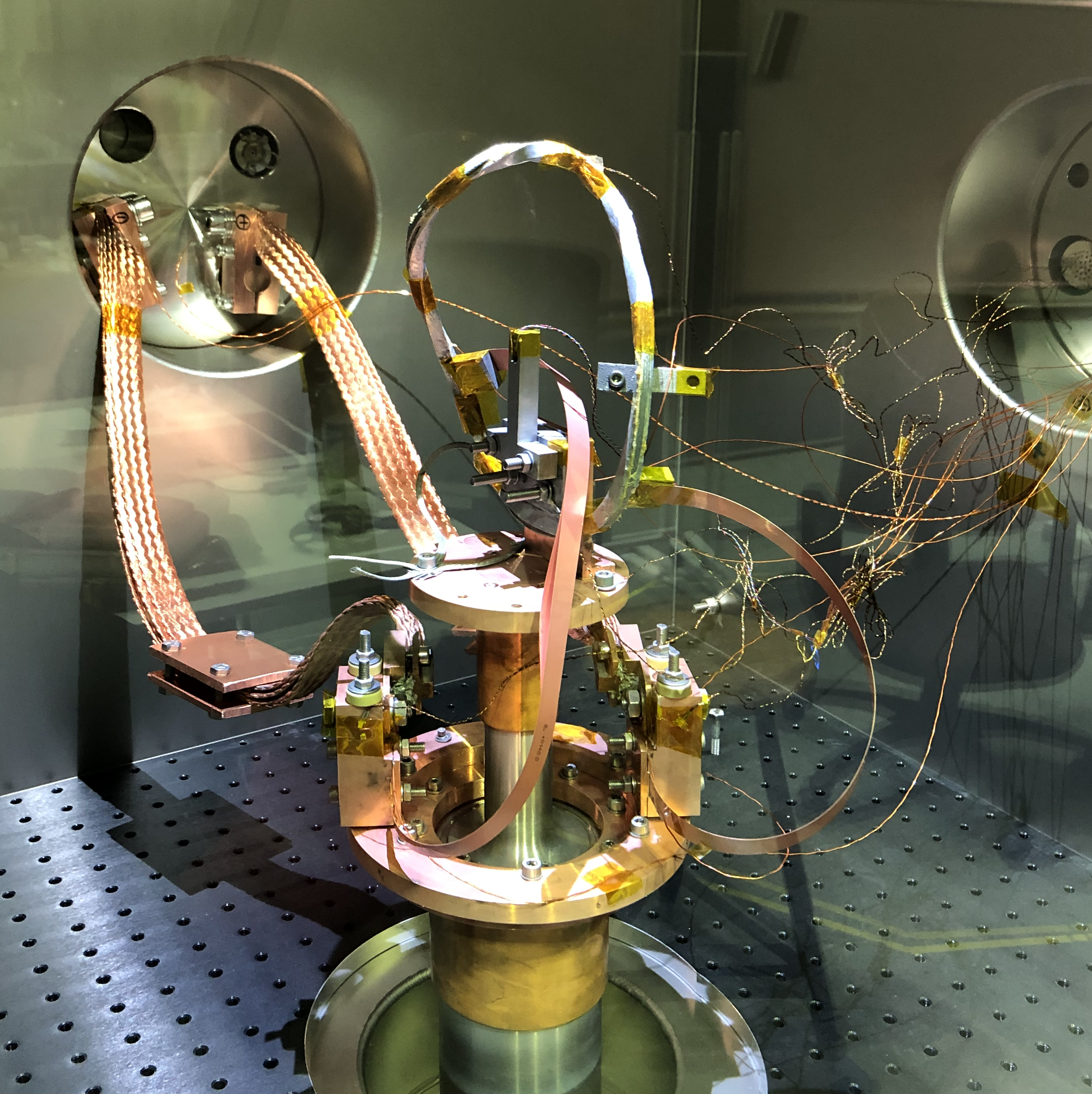}
    \caption{The coil mounted onto a Leybold 130 MD cryocooler. 60 mm$^2$ thick, braided copper leads supply the current to the first stage at a temperature of $\approx60\,K$ from which 12\,mm ReBCO leads connect to the winding pack of the coil that is cooled down by the second stage of the cryocooler. This setup resembles the operating setup planned for EPOS.}
    \label{fig:enter-label}
\end{figure}

In a second attempt, there was no solder used between the layers. 21 layers were wound onto the coil which now was significantly easier due to the extensive polishing. For good thermal coupling to the frame, we filled the trench with Apiezon N grease and wound a 3\,mm wide copper braid around the outide of the winding pack. The magnetic field measured in liquid nitrogen is depicted in figure \ref{fig:bfield}. It follows the expectation for the coil and reaches up to 7.89\,mT. This suggests - together with the low resistance $R_{wp}=(703\pm13)\,n\Omega$ found along the winding pack - that we manufactured the coil without damage to the superconducting layer of the ReBCO tape. For the contact resistance across the solder contacts between the 12\,mm current leads and the 3\,mm windings, we found $R=(2.026\pm0.045)\mu\Omega$ for one of the contacts. A faulty voltage probe prevented the measurement of the second contact. This is comparable to the resistance found in \cite{huslage_winding_2024} for a coil without defects.


\begin{figure}[!t]
\centering
\includegraphics[width=\columnwidth]{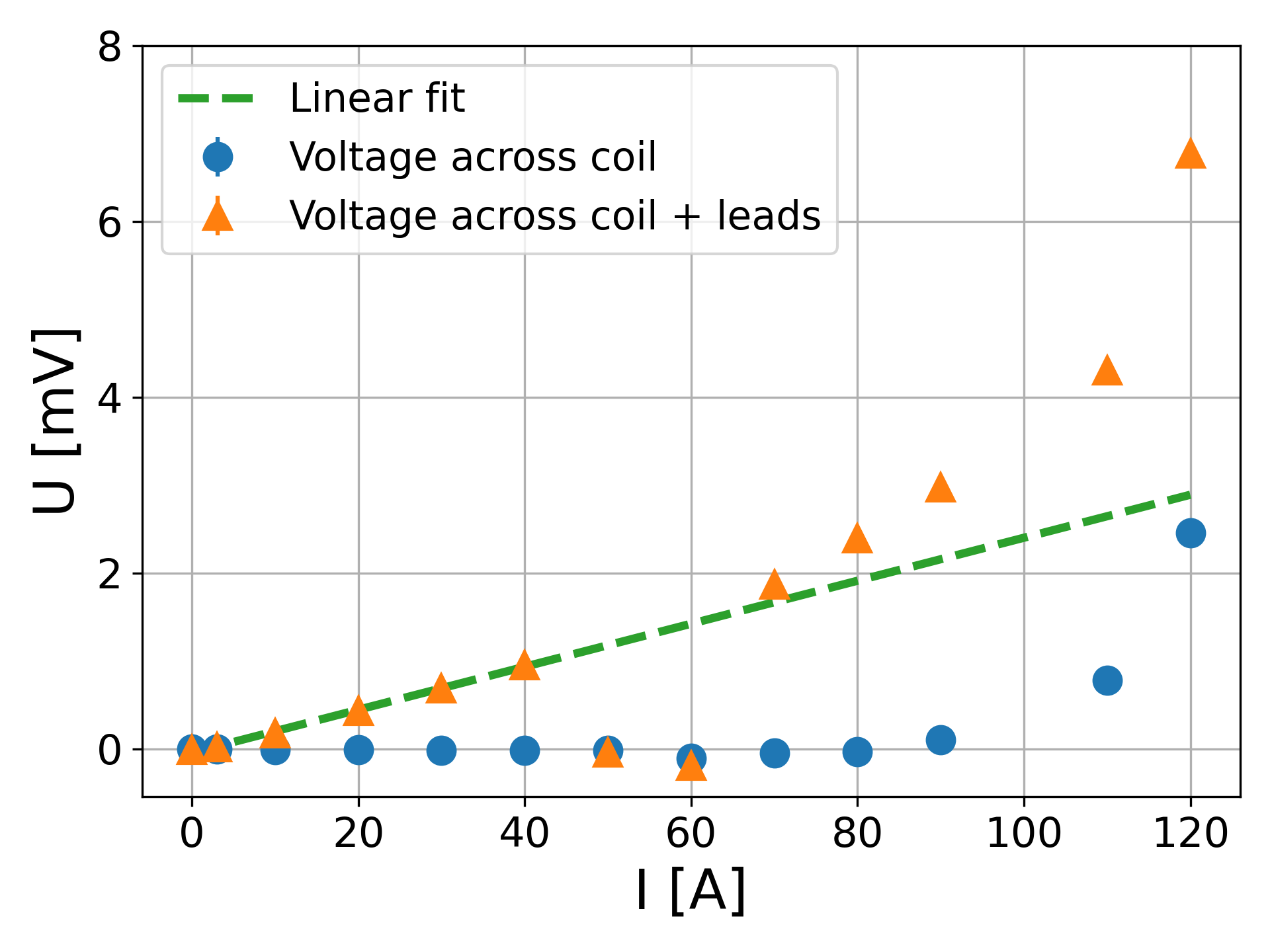}%
\caption{Four-point measurement of the voltage across superconducting winding pack (blue) and the 12\,mm current leads (orange) cooled in a cryocooler. The linear fit to the voltage across the current leads gives a contact resistance of $R=\left(2.25\pm0.13\right)\mu\Omega$. Its earlier deviation from linearity suggests that the quench originates from the current leads (and most probably from the heat induced by the contact resistances).}
\label{fig:voltage}
\end{figure}

In order to test the coil close to the EPOS operating conditions, we mounted the coil onto a cryocooler inside a vacuum chamber using the plate that was added to the winding frame. The cryo cooling has two cooling stages. Reaching the EPOS operating temperature proofed to be challenging as the thin, 3D-printed winding pack showed poor thermal conductivity leading to large temperature gradients between top and bottom of the coil. The target operating condition could only be achieved with the introduction of a heat shield and a thick copper braid connecting the top of the coil to the base of the cold head. We did not use those measures when energizing the coil on the cold head as they were in the way of the current leads.

The coil was energized up to 120\,A in the vacuum chamber being cooled with a Leybold 130 MD cryocooler. An axial, cryogenic hall probe (HGCA-3020) was placed close to the center of the coil. The measured magnetic field shown in figure \ref{fig:bfield} follows the theoretical prediction. The coil achieved a peak field of 21.2\,mT. This agrees well with the magnet field calculated from inserting the optimized coil spline into the Biot-Savart law. 

The high inductance of NI coils can lead to potentially very high charge and discharge times. The characteristic time scale of the charge and discharge process is given by $\tau_C=L/R$, where $L\sim N^2$ is the inductance of the coil which is proportional to the square of the number of windings. $R$ is the resistance of the coil. In the event of a sudden discharge (switching off the power supply at $t=0$), the magnetic field of the coil after a time $t$ is expected to follow an exponential decay
\begin{equation}
    B = B_0e^{-t/\tau_C}.
\end{equation}

\begin{figure}[!t]
\centering
\includegraphics[width=\columnwidth]{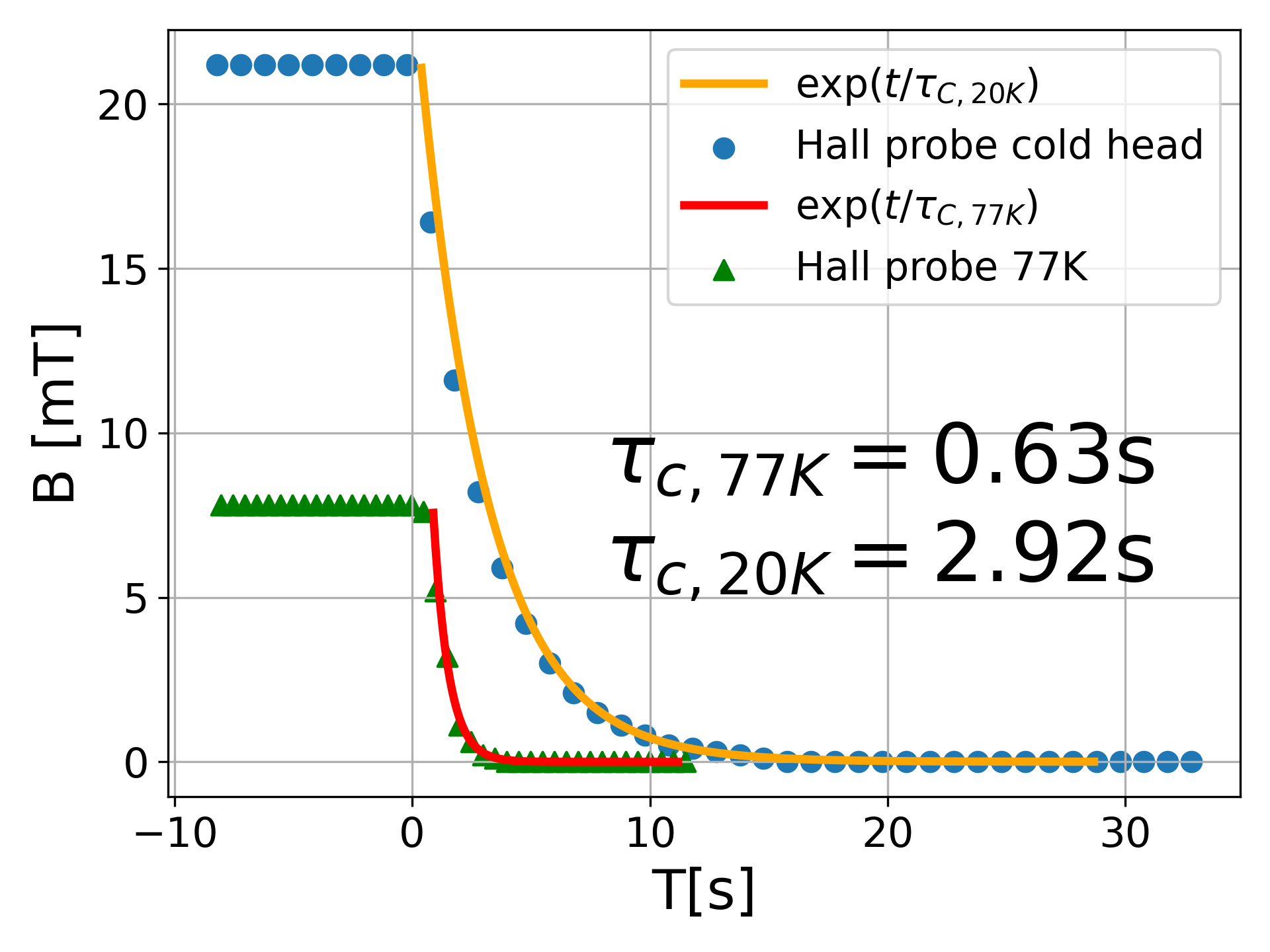}%
\caption{Discharge times $\tau_c = L/R$ of the test coil in liquid nitrogen and cooled with a cryocooler. We obtain a estimate of the winding pack temperature on the cryocooler of $41\,K$ from the ratio of the two discharge times and the corresponding changes in copper resistance.}
\label{fig:l_over_r_time}
\end{figure}

$B_0$ is the magnetic field at $t=0$. This time $\tau_C$ was found to be 0.63\,s in liquid nitrogen and 2.92\,s on the cold head, as shown in Fig. \ref{fig:l_over_r_time}. We can use the ratio of the discharge times to obtain the temperature of the superconducting winding pack, assuming the current in the coil is dissipated through the copper surround layers. As $\tau_{c, LN2}/\tau_{c, CC} = R_{CC}(T)/R_{LN2}(77\,K)$ . We find an estimated temperature of 41\,K in the winding pack
If we account for the difference in copper resistance between 41\,K and 20\,K and scale up the number of windings, this would lead to an estimated charge/discharge time for EPOS of 
\begin{equation}
    \tau_{C, EPOS} = \left(\frac{21}{150*22}\right)^2\tau_{C, cc} \approx 20\,h
\end{equation}
This is a large, but tolerable time for the ramp-up of the field for the stellarator, which is meant to operate in steady state.

\section{Summary and outlook}\label{sec:summary}
In this paper, we demonstrated the operation of a small-scale ($d\approx0.2\,$m), non-insulated ReBCO coil wound onto a 3D-printed winding frame both in liquid nitrogen and cooled with a cryocooler. 

We optimized the orientation of the superconducting winding pack to avoid damage to the tape from torsional and hard-way bending strains and 3D-printed the winding frame out of AlSi10MG using selective laser melting. The manufacturing deviations obtained by a 3D scan was found to be 0.3\,mm. This is in line with the manufacturer's specifications. The 3D printed winding frame showed a high surface roughness which significantly impeded the winding process. The first attempt at winding showed a lower than expected magnetic field due to defects. Only after time consuming post processing of the winding frame using a hand held micro tool, could the coil be wound without damage to the tape.

We energized the coil up to 40\,A supply current in liquid nitrogen and achieved a magnetic field of 7.89\,mT. Cooled by a cryocooler, the coil achieved a field of 21.2\,mT at a supply current of 120\,A. This agrees well with the theoretical expectation. 

The resistance measurements along the superconducting windings suggest, that we found the tape around the winding frame without introducing defects to the ReBCO tape. For the contact resistance, we found values of $R=\left(2.25\pm0.13\right)\mu\Omega$. At high operating currents in the cold head experiments, the heat load from Ohmic losses at the solder contacts between 12\,mm wide superconducting leads and 3\,mm wide windings, likely lead to the quench of the coil.

The charge/discharge time of the coil $\tau_C=L/R$ was found to be 0.63\,s in liquid nitrogen and 2.92\,s on the cryocooler. The differing charging time suggests a temperature of 41\,K achieved on the cold head. Scaled up to the full EPOS stellarator, we expect a charging time of approximately one day.

A few lessons were learned from the construction of this test coil: The high surface roughness of the 3D-print significantly hindered the winding of the coil. An easier winding process could be achieved by significant post processing or a wider trench, which would in turn result in increased manufacturing deviation. Furthermore, the dimness of the surface seems to complicate the cooling due to decreased reflection of black body radiation. 

Future work aims to get closer to the EPOS operating conditions. We will construct further test coils with more turns and several pancakes to achieve higher magnetic fields. We will also analyze the manufacturing deviations on this and other coil frames and present the results in a future publication.

\section*{Acknowledgments}
P.H. thanks Robert Luerbke, Adam Deller, Alex Card, Raphael Unterrainer, Melanie Russo and the entire CSX team for useful discussions. Furthermore, we want to thank 3D-activation for manufacturing the coil frame and providing consultation during the process. This work has been supported by the Helmholtz Association (VH-NG-1430).
 
\bibliographystyle{unsrt}

\bibliography{sn5_bio.bib}

\newpage

\vspace{11pt}

\vfill

\end{document}